\documentclass[10pt]{iopart}
\usepackage{graphicx,float,epsfig,wrapfig}
\usepackage[nooneline,scriptsize,bf]{caption2}

\begin{document}

\title[Building a 3.5 m prototype interferometer ...]{Building a 3.5 m prototype interferometer for\\
the Q \& A vacuum birefringence experiment\\
and high precision ellipsometry}

\author[J-S Wu, W-T Ni, and S-J Chen]{Jeah-Sheng Wu\footnote[3]{
Present Address:  Center for Meas. Standards, ITRI,
Hsinchu, Taiwan, ROC (jswu@itri.org.tw)},
 Wei-Tou Ni and Sheng-Jui Chen}

\address{Center for Gravitation and Cosmology, Department of Physics,\\
National Tsing Hua University,
Hsinchu, Taiwan 30055, ROC}

\begin{abstract}
We have built and tested a 3.5 m high-finesse Fabry-Perot prototype inteferometer
with a precision ellipsometer for the QED test and axion search (Q \& A) experiment.
We use X-pendulum-double-pendulum suspension designs and automatic control schemes
developed by the gravitational-wave detection community.
Verdet constant and Cotton-Mouton constant of the air are measured as a test.
Double modulation with polarization modulation 100 Hz and
magnetic-field modulation 0.05 Hz gives $10^{-7}$ rad phase noise
for  a 44-minute integration.
\end{abstract}

\pacs{04.80.-y, 12.20.-m, 14.80.Mz, 07.60.Ly, 07.60.Fs, 33.55.Ad}



\section{Introduction}
Quantum Electrodynamics (QED) predicts that in a background electromagnetic field, vacuum is refractive
and birefringent.  The  refractive
indices in a transverse external  magnetic   field
$\vec{B}^{\rm ext}$ are
$n_{\parallel}=1+3.5
{{\alpha (B^{\rm ext})}^{2}}/(45\pi B_{\rm c}^2)$ and
$n_{\perp}=1+2\alpha {(B^{\rm ext})}^{2}/(45\pi B_{\rm c}^2)$
for linearly polarized  lights
whose polarizations are parallel and orthogonal to the magnetic field with
$B_{\rm c}={{m^{2}c^{3}}/{e\hbar}}=4.4\times 10^{9}$ ${\rm T}$.
For a transverse magnetic field
(dipole field) of 2.5 T, $\Delta n = n_{\parallel}-n_{\perp} = 2.5 \times 10^{-23}$.  This
birefringence is measurable using double-modulation ultra-high sensitive interferometer techniques.
In typical invisible axion models, axion-photon coupling induces both ellipticity and
polarization-rotation for light propagation in a magnetic field; these effects are about 7
orders smaller in magnitude compared to the QED birefringence.

In 1994, we proposed the Q \& A experiment to measure the vacuum birefringence and
the axion-photon coupling [1], and began to construct and test a 3.5 m
high-finesse Fabry-Perot prototype inteferometer.
In [1], we presented the motivation and background of this experiment in detail.
In [2], we present methods of improvement and make a comparison of Q \& A experiment,
PVLAS experiment [3], and BMV experiment [4].
Here we present the experimental setup, and test-measurement results.


\section{Experimental setup of the 3.5 m prototype interferometer}
The 3.5 m prototype interferometer (figure 1) is formed using a high finesse Fabry-Perot interferometer together with a high precision ellipsometer.
The two high-reflectivity mirrors of the 3.5 m prototype interferometer are suspended separately
from two X-pendulum-double-pendulum suspensions (basically the design of ICRR, University of Tokyo [5])
mounted on two isolated tables fixed to ground using bellows inside two vacuum chambers. The characteristics
of these vibration isolation sub-systems are to be discussed in [2]. Other sub-systems are described below.

\begin{figure}[h]
\includegraphics[width=\linewidth,height=4.5cm]{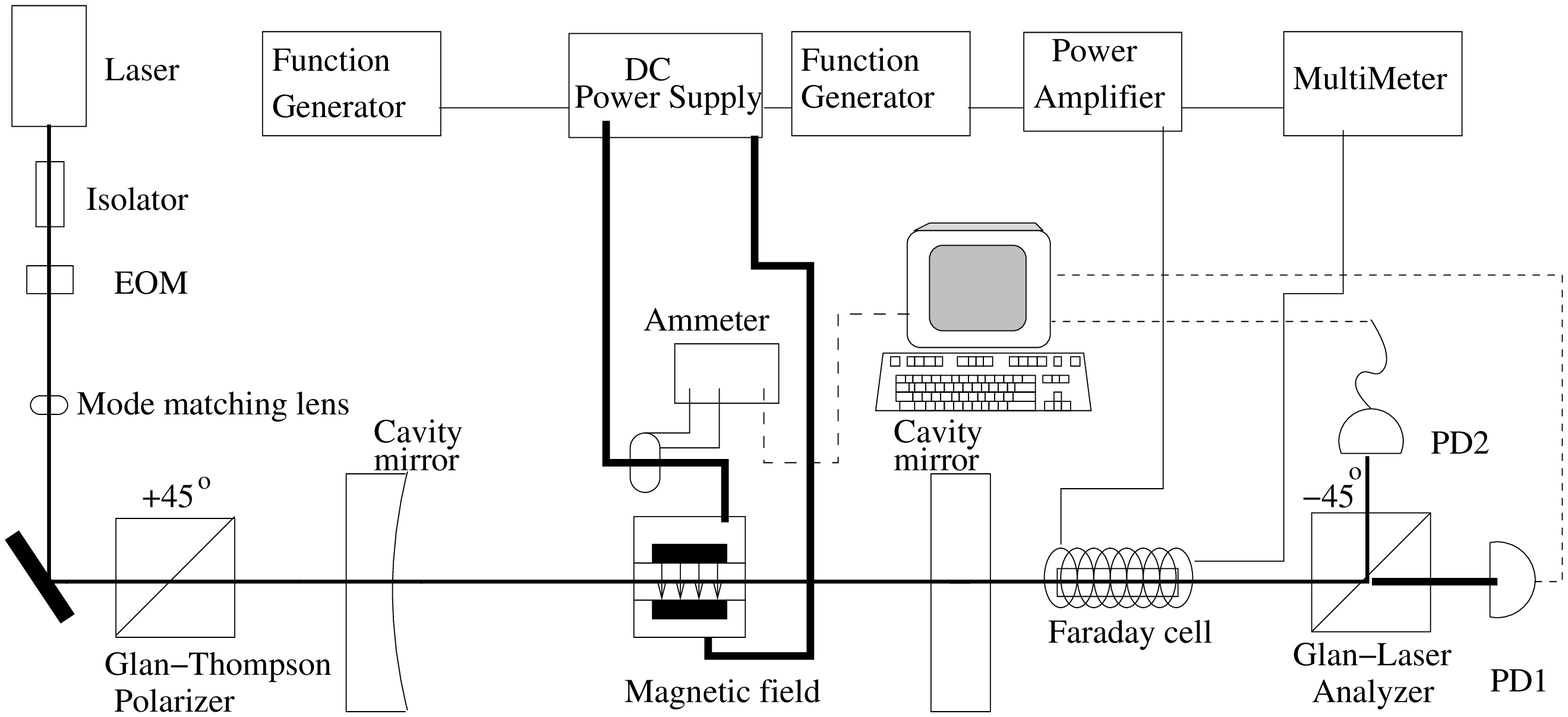}
\begin{center}
\mbox{\scriptsize {\bf Figure 1.} Schematic diagram of the 3.5 m prototype interferometer}
\end{center}
\end{figure}

{\it Laser, cavity and finesse measurement} ---
We use a 1 W diode-pumped 1064 nm CW Nd:YAG laser made by
Laser Zentrum Hannover as light source.
The laser frequency can be tuned using thermal control and PZT control.
\begin{wrapfigure}[16]{r}{.55\linewidth}
\includegraphics[width=\linewidth]{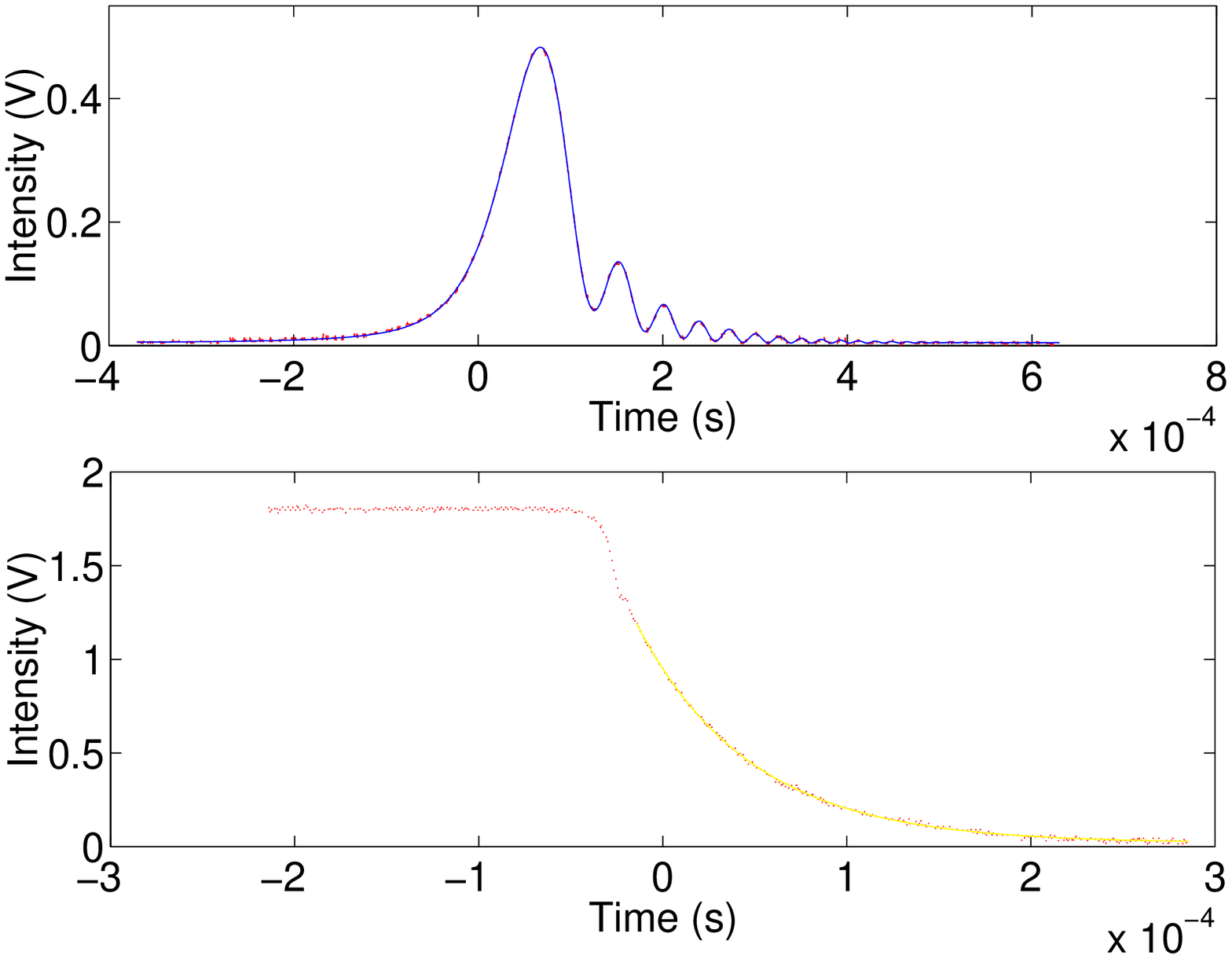}
\mbox{\scriptsize ~~~~~~~~~~~~~ {\bf Figure 2.} Two finesse measurements.}
\end{wrapfigure}
We thermal-stabilize the laser at 25 $^\circ$C and use Pound-Drever scheme
to lock the laser frequency with the resonant frequency of the 3.5 m interferometer.
The high-frequency ($>$ 30 Hz) part of the phase error from the photodetector is fed to the PZT to frequency lock the laser
to the cavity resonant frequency;
this feedback circuit is a three stage amplification with lead-leg compensation.
Lower frequency part ($<$ 30 Hz) of the phase error goes to a DSP based digital control unit
to control the cavity length of the interferometer.
The laser goes through an isolator, a 882 MHz electo-optic modulator and a mode-matching lens to the 3.5 m cavity.
The modulation frequency was chosen as the free spectral range of a 0.17 m mode cleaner
which was built but not used.
We measure the finesse of the 3.5 m cavity in 3 ways while the laser was locked to the resonant cavity.
Two methods [6,7] are to analyze the transmission intensity profile,
as shown in upper part of figure 2.
The third approach is to measure the life time of intra-cavity photons.
Lower part of figure 2 shows the decay of transmision intensity when the power of the input light was
suddenly turned off.
All three methods give finesse 17,000 to within 1 \%.
This finesse enhances the cavity birefringence measurement by
10,800 ($2F/\pi$) times.

{\it Polarizing optics and ellipsometry} ---
A Glan-Thompson polarizer with measured extinction ratio $2.60\times 10^{-7}$ is placed before the 3.5 m cavity.
A Glan-Laser polarizer with measured extinction ratio $9.36\times 10^{-7}$ is placed behind the cavity as analyzer.
A Faraday glass made of terbium-boron-alumina-silicate oxide with terbium oxide greater than 50\% by weight
(Kigre model M-18) is placed before the analyzer for the purpose of signal modulation.
This Faraday glass is housed in a Teflon mount wound around
with 2500 turns of 0.2 mm enamel-insulated wire.
Modulation response of the Faraday glass was measured to be $\eta=0.019$ rad A$^{-1}$.
The polarizer, the analyzer and the Faraday glass
form a modulated Malus ellipsometer to measure the intra-cavity polariztion change.  Malus law
$P = P_0 (\epsilon ^2 + \psi ^2)$ gives the polarization rotation angle $\psi$ from the power received in the photodetector.
$\epsilon ^2$ is the extinction ratio of the Malus ellipsometer.

{\it Magnet} --- We use a switching dipole magnet
with a 25.4 mm borehole to induce the intra-cavity birefringence of the 3.5 m prototype interferometer.
This magnet can generate up to 1.2 T transverse magnetic field with an effective magnetic length 0.2 m.
A vacuum tube of ID/OD 21.4 mm/24.6 mm goes through the borehole of the magnet to connect the
two mirror-hanging vacuum chambers.

\section{Verdet constant of the air}
\begin{wrapfigure}[16]{r}{0.45\linewidth}
\includegraphics[width=\linewidth]{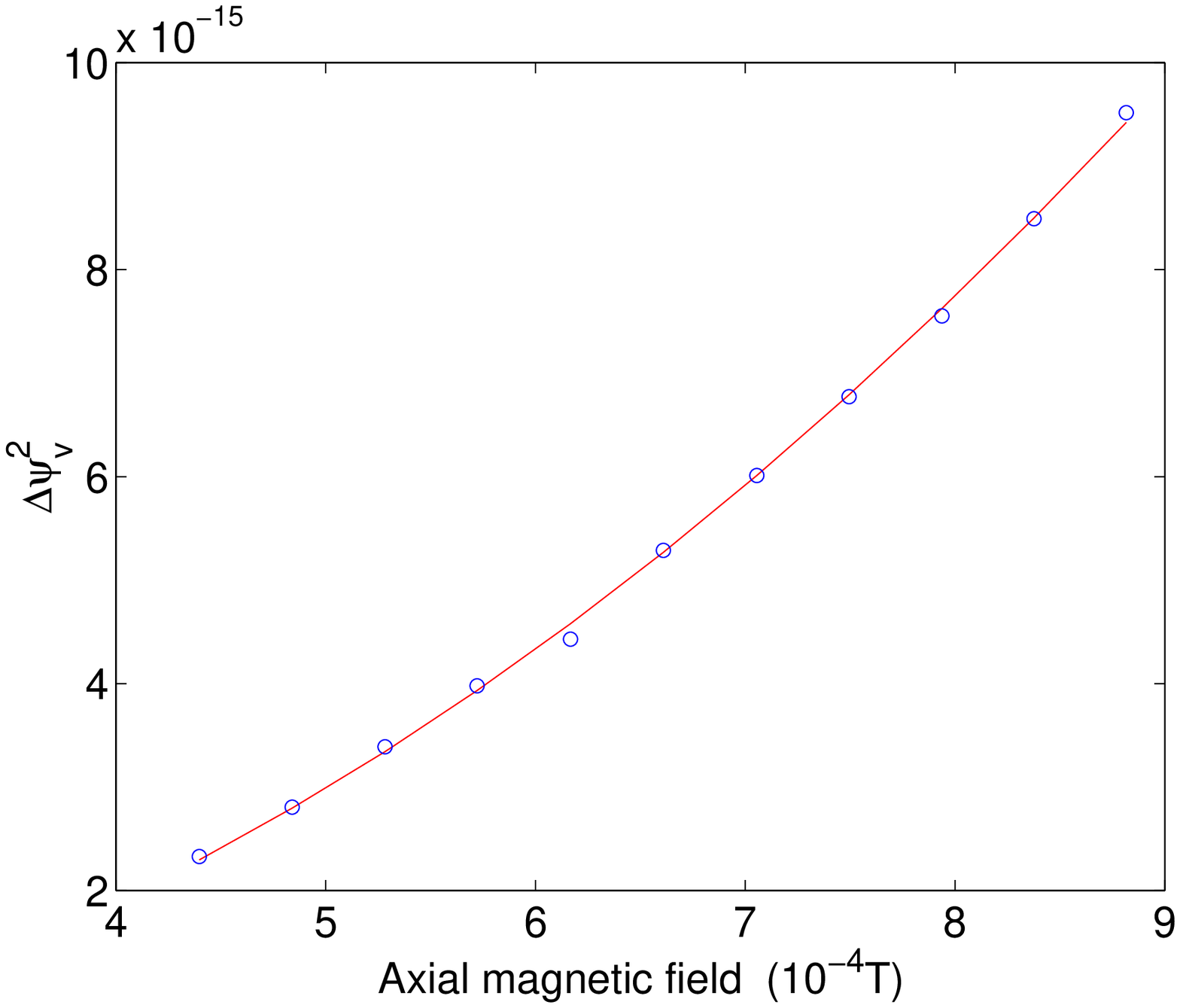}
\mbox{\scriptsize {\bf Figure 3.} {$\Delta\psi_v^2$} vs. B for measurement of the}
\mbox{\scriptsize Verdet constant of air. }
\end{wrapfigure}
Faraday effect is a property of transparent
substance in a magnetic field
which induces a rotation of the plane of polarization with distance
for light propogated along the magnetic field.
For dilute material like gas,
the Faraday rotation is far smaller compared to substance like water or glass.
We measure the Faraday rotation of air
(polarization rotation angle $\psi_v=C_{v}B_0L_{eff}$)
and determine the Verdet constant $C_v$ using the 3.5 m prototype interferometer.
We use a 0.4 m long home-made solenoid to apply a 100 Hz axial magnetic field $B=B_0cos(\omega t)$,
to the 3.5 m cavity.  From the response of the photodetector,
we obtain the polarization signal
$\rho \equiv P/P_0=\epsilon^2+(4\mathcal{F}^2)/(\pi^2)[(\Delta\psi_{v}^2)/(2)+(\Delta\psi_v^2)/(2)(cos(2\omega t))]$.
When the amplitude of applied magnetic field is varied from $4.4\times 10^{-4}$ T to $8.8\times 10^{-4}$ T,
the 200 Hz demodulated signals gives $\Delta\psi_v^2$ shown as ordinates in Fig. 3.
A quadratic fit to the magnetic field determines the Verdet constant to be
$C_v=(3.91\pm 0.02) \times 10^{-4}$ rad T$^{-1}$m$^{-1}$
at 25.5$^\circ$C and 1 atm for $\lambda = 1064$ nm.

\section{Cotton-Mouton effect}

Cotton-Mouton effect is quadratic to transverse magnetic field: $\psi_{CM}=\pi C_{CM} B^2 L_{eff}$.
For measuring the Cotton-Mouton effect,
we replace the axial field by the transverse field $B=B_0+B_m cos(\omega_m t)$ of the switching magnet and
follow a similar procedure as described above.  Since the alignment of the
magnet axis and optical axis is not perfect, there is a small axial magnetic field $B_{ax}=kB$
to induce a Faraday rotation.
\begin{table}
\begin{center}
\mbox{\scriptsize {\bf Table 1.} Data of $B_0$, $B_m$, and $\rho_{\omega_m}$ for 7 experimental runs of measuring Cotton-Mouton effect.}
\begin{tabular}{llllllll}
\hline
$B_0 (10^{-3}$ T)&64.56&87.60&111.84&134.64&156.48&180.24&204.72\\
$B_m (10^{-3}$ T)&46.56&58.56&70.32&82.08&94.08&105.84&117.60\\
$\rho_{\omega_m}(10^{-7})$&1.22& 1.94& 2.29& 3.31& 3.90& 5.18& 6.26\\
\hline
\end{tabular}
\end{center}
\end{table}
The detected polarization signal $\rho$
contains both Faraday rotation and Cotton-Mouton birefringence:
$\rho(t)=\epsilon^2+\frac{4\mathcal{F}^2}{\pi^2}(\pi C_{CM}(B_0+ B_m cos(\omega_m t))^2L_{eff}
        +kC_v(B_0+ B_m cos(\omega_m t))L_{eff}+\psi_c)^2$.
Where $\psi_c$ is the birefringence of mirror coating.
Demodulating $\rho(t)$ at $\omega_m$ we obtain
\begin{eqnarray*}
\rho_{\omega_m}&=&\epsilon^2+\frac{2\mathcal{F}^2}{\pi^2}B_mL_{eff}(kC_v+2B_0C_{CM}\pi)\\
               & & (B_0C_vkL_{eff}+4B_0^2C_{CM}L_{eff}\pi+3B_m^2C_{CM}L_{eff}\pi+4\psi_c).
\end{eqnarray*}
We modulate the magnetic field, and demodulate the detected signal at $\omega_m=0.05$ Hz to give
$\rho_{\omega_m}$.
By varying $B_0$ and $B_m$, we obtain 7 sets of data (Table I).
Fitting the 7 data points of $\rho_{\omega_m}$ using simplex methods,
build-in Matlab function of nonlinear fitting,
we obtain $C_{CM}=(5.50\pm 0.48)\times 10^{-7}$ rad T$^{-2}$ m$^{-1}$ and
$k = (9.58\pm 1.29)\times 10^{-4}$.

\section{Sensitivity curve of prototype interferometer}
\begin{wrapfigure}[12]{r}{0.6\linewidth}
\includegraphics[width=\linewidth,height=3.5cm]{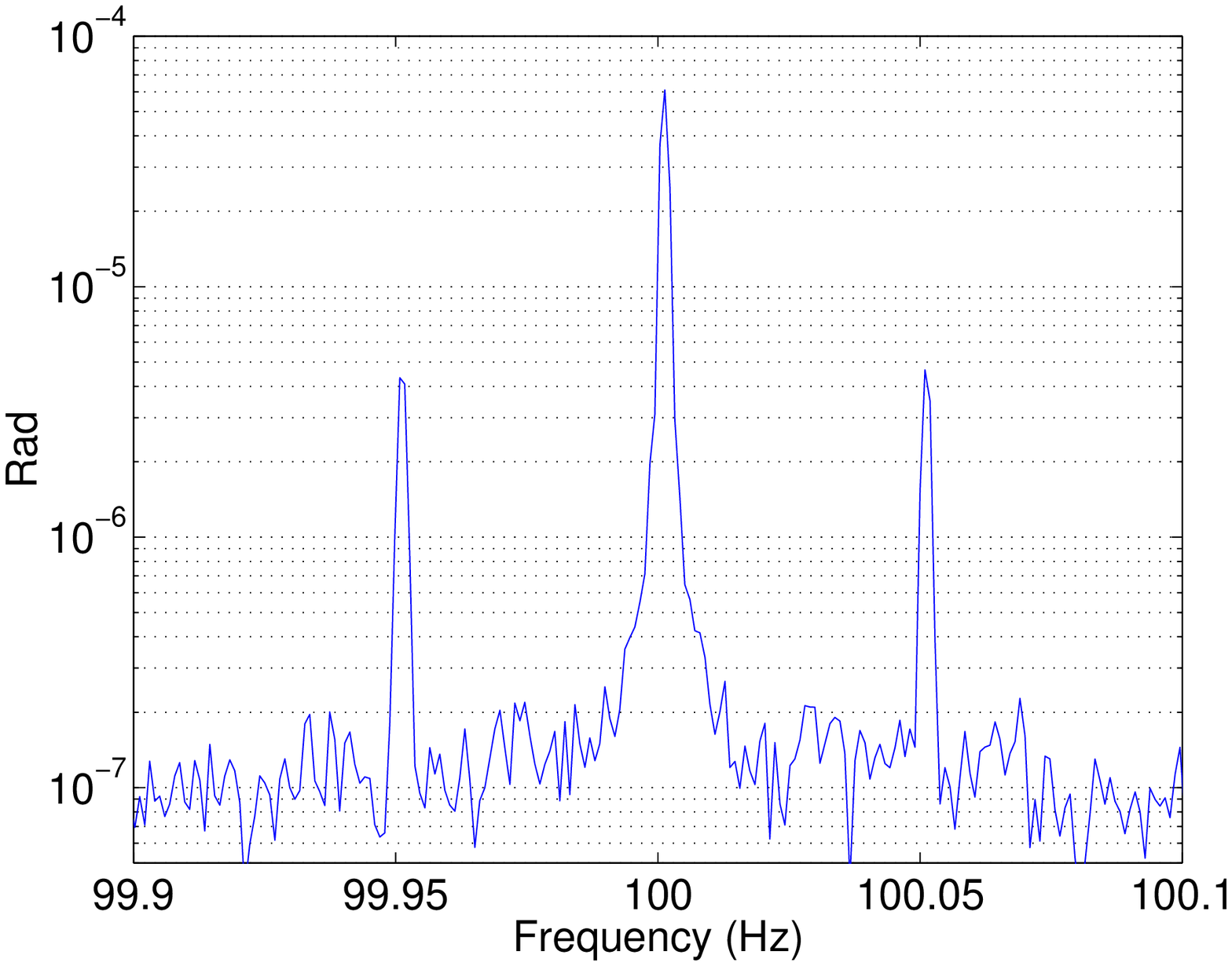}
\mbox{\scriptsize {\bf Figure 4.} Noise floor of the 3.5 m prototype interferometer}
\mbox{\scriptsize  around the double-modulation frequencies.}
\end{wrapfigure}
  The sensitivity for detecting polarization rotation of prototype interferometer
 can be estimated by Fourier analyzing the double modulated signal (polarization modulation at 100 Hz using Faraday cell
and magnetic field moduation at 0.05 Hz using switching dipole magnet)
to obtain power sprectrum using Welch's average and Hanning window.
The two side peaks are due to the coupling between the static birefringence of the Faraday cell and Cotton-Mouton effect of
the air.
The noise floor near modulated frequency 100 Hz $\pm$ 0.05 Hz
is around $10^{-7}$ rad ($5\times 10^{-6}$ rad Hz$^{-1/2}$) as shown in figure 4 for a
44-minute integration.
For measuring vacuum birefringence, a quarter-wave plate needs to be introduced between the output cavity mirror and
Faraday cell. In [2], we present 
the schemes and current status of the 2nd and 3rd phases of the Q \& A experiment.

We thank the National Science Council for supporting this research in part.

\References
\bibitem[1] {Ni} Ni W-T 1998
{\it Frontier Tests of QED and Physics of the Vacuum}
ed. E Zavattini et al
(Sofia: Heron Press) p~83; and references therein
\bibitem[2] {Chen} Chen S-J, Mei H-H, Ni W-T and Wu J-S 2003
Improving the Ellipticity Detection Sensitivity for the Q \& A Vacuum Birefringence Experiment Talk presented at
{\it 5th Amaldi Conf. (Tirrenia, July 2003)}
\bibitem[3] {}Brandi F \etal 2001
{\it \NIM}, {\bf 461} 329
\bibitem[4] {}Askenazy S \etal 2001
{\it Quantum Elec. and Phys. of the Vacuum}
ed. G Cantatore (AIP) p~115
\bibitem[5] {xpen}Barton M \etal 1999
\RSI {\bf 70} 2150; Tatsumi D \etal 1999 {\it ibid} 1561
\bibitem[6] {swpfp1} Li Z, Stedman G E and Bilger H R 1993
{\it Opt. Comm.} {\bf 100} 240
\bibitem[7] {mfp} Vallet M \etal 2001
{\it Opt. Comm.} {\bf 168} 423; and references therein
\endrefs
\end{document}